\documentclass[12pt,preprint]{aastex} 
\tighten 

\begin{document}

\received{6 July 2004}
\accepted{1 September 2004}

\title{Effects of Metallicity on the Rotational Velocities of Massive Stars}

\author{Laura R. Penny, Amanda J. Sprague, George Seago} 
\affil{Department of Physics and Astronomy \\
       The College of Charleston \\
       Charleston, SC 29424 \\
       pennyl@cofc.edu, ajspragu@edisto.cofc.edu, gseago@edisto.cofc.edu}
\author{Douglas R. Gies}   
\affil{ Center for High Angular Resolution Astronomy and \\ 
       Department of Physics and Astronomy,\\
       Georgia State University, Atlanta, GA 30303-3083 \\ 
       gies@chara.gsu.edu}



\begin{abstract} 

Recent theoretical predictions for low metallicity massive stars predict 
that these stars should have drastically reduced equatorial 
winds (mass loss) while on the main sequence, and as such should retain most of their 
angular momentum.   Observations of both the Be/(B+Be) ratio and the 
blue-to-red supergiant ratio appear to have a metallicity dependence that 
may be caused by high rotational velocities.  We have analyzed 39 
archival {\it Hubble Space Telescope Imaging 
Spectrograph (STIS)}, high resolution, ultraviolet spectra
of O-type stars in the Magellanic Clouds to determine their
projected rotational velocities $V\sin i$.  Our methodology is based on a 
previous study of the projected rotational velocities of Galactic O-type stars using
{\it International Ultraviolet Explorer (IUE)} Short Wavelength Prime (SWP) Camera high 
dispersion spectra, which resulted in a catalog of $V \sin i$ values for 177 O stars.
Here we present complementary $V\sin i$ values for 
21 Large Magellanic Cloud and 22 Small Magellanic Cloud O-type stars
based on STIS and {\it IUE} UV spectroscopy.  The distribution 
of $V \sin i$ values for O type stars in the Magellanic Clouds 
is compared to that of Galactic O type stars. 
Despite the theoretical predictions and indirect observational evidence for high rotation, 
the O type stars in the Magellanic Clouds do not appear to rotate faster 
than their Galactic counterparts.
\end{abstract}

\keywords{Stars: spectroscopic ---
 stars: early-type --- 
 stars: rotation ---
 Magellanic Clouds ---
 ultraviolet: stars}

\pagebreak

\section{Introduction} 

What effect does metallicity have on the mass loss and angular momentum loss of massive stars?  
There is a large amount of secondary evidence, some of which is summarized below, 
suggesting that low metallicity should result in smaller angular momentum loss 
and subsequently higher rotational velocities for these type stars.  However, 
there is an unfortunate lack of direct measurements of the equatorial rotational velocity $V_{rot}$, 
or even of the projected rotational velocity $V \sin i$ for massive stars in a 
low $Z$ environment.  The Large and Small Magellanic Clouds (LMC \& SMC) are nearby, contain many 
young massive objects, and have low $Z$ values of 0.007 and 0.004, respectively.  As such 
they present an ideal laboratory to test the dependence of massive star rotation on metallicity. 

Maeder, Grebel, \& Mermilliod (1999) found a significant increase in the 
number ratio of Be/(B+Be) stars with decreasing
metallicity among Galactic and MC clusters 
(see, however, the counterarguments presented by Keller et al.\ 1999). 
The Be phenomenon is closely linked to rapid rotation (Porter \& Rivinius 2003).  
To explain this trend in Be star number ratio, the authors raise the possibility 
that stars forming in a low metallicity environment would
have higher initial rotational velocities.  
Although no numerical models of the formation of massive 
stars at low $Z$ have been produced, the authors tentatively present 
some possible origins of these higher initial
rotational velocities.  A lower $Z$ environment would have lower dust content 
and fewer metallic ions present in star
forming regions so that the magnetic field of the contracting central mass 
is less coupled to the surrounding region. 
Also the lifetime of the accretion disk is likely related to its opacity, 
which will decrease with lower metallicity. 
These factors would result in less angular momentum loss during star formation, 
leading to higher initial rotational
velocities.  The possibility of faster initial rotation at low metallicity 
resulting in more rotational mixing on the
main sequence has also been suggested to resolve the nitrogen enrichment seen 
in SMC B- and A-supergiants (Venn et al.\ 1998).  Maeder et al. (1999) state 
``Of course, direct observations of $V\sin i$ in LMC and SMC clusters are very much needed
in order to substantiate the above results.''

Observations of the number ratio of blue to red supergiants in nearby galaxies also 
show a sharp dependence upon metallicity (Humphreys \& McElroy 1984). 
Stellar interior models which include rotation can account for this long-standing problem
(Maeder \& Meynet 2001).  The primary cause of this is the mild mixing just outside 
the core of a rotating star during the main sequence.  This increases the amount of 
helium near and above the H-shell.  The net effect of the larger He
core and the mild main sequence mixing is to decrease the importance 
of the convective zone above the H-burning shell in
the post-main sequence evolution.  Because of the small polytropic index 
in the convective zone, the larger the size of the zone the
smaller the radius of the star.  Conversely, if its contribution is small, 
the overall size of the star increases,
leading to a redward position in the Hertzsprung-Russell (HR) diagram. 

Another key prediction of these new models is shown by comparing 
Figure 10 of Meynet \& Maeder (2000)  and Figure 3 of Maeder \& Meynet
(2001).  These are plots of the evolution of surface rotational velocities 
for massive stars in both Galactic and SMC metallicity, respectively.  
At low $Z$, the surface rotational velocities will remain almost constant 
during the main sequence evolution, contrary to that of massive stars at solar
metallicity where the rotational velocities are predicted to rapidly decrease.  
Theoretically this results from both lower equatorial mass loss 
and the outward transport of angular momentum from the interior of 
the star by meridional circulation.  Models of massive stars at both metallicities
show a drastic decrease in rotational velocity after the terminal age main sequence. 

The idea that a low metallicity environment would create massive, 
rapidly rotating, helium stars has even been raised by
theorists as a possible source of gamma-ray burst progenitors 
(MacFadyen \& Woosley 1999).  Here the outer layers of the
star, and, in at least some cases, the mantle have too much angular momentum 
to fall freely inside the last stable orbit
during core collapse. An accretion disk forms where the dissipation of 
rotational and gravitational energy will give
rise to some sort of mass ejection and electromagnetic radiation, a ``hypernova.'' 
This results in the prompt formation of a black hole through hyperaccretion. 

\section{Projected Rotational Velocities from UV lines} 

Stellar rotation is the dominant process in shaping the
photospheric lines of O stars. The vast number of lines 
present in the ultraviolet spectra of O-type
stars suggests that accurate projected rotational velocities 
could be found by studying these objects in the
ultraviolet.  These lines arise from high excitation transitions 
from deeper in the photosphere, and they are less subject to
wind variability than optical lines.  Also they are less likely 
to be contaminated by weak emission from circumstellar
and nebular gas.  The standard method of individual profile 
fitting in the UV is difficult because of line blending.  
A powerful alternative involves cross-correlation of a 
narrow-lined star with a test star, which will result in a
cross-correlation function (ccf) that represents a 
``superline'' of the test star.  This technique actually increases the
S/N of the observed spectrum by in effect combining all the 
UV lines into one. The ``superline'' will have an observed 
width related to the line width of the test star and that of
the narrow-lined star (Figure 1).  Fitting this ``superline" is 
the numerical equivalent of fitting each
UV absorption feature and combining the resulting measurements. 
Our previously derived $V \sin i$ values for 177
Galactic O stars were calculated by calibrating the widths of ccfs 
derived from {\it IUE} SWP high dispersion 
spectra as a function of published $V \sin i$ values from 
Conti \& Ebbets (1977) in a common sample of
objects (Penny 1996 = P96).  Conti \& Ebbets (1977) fit artificially 
broadened model line profiles (from the non-LTE model atmospheres of 
Auer \& Mihalas 1972) of four visual absorption lines to determine
their $V \sin i$ values.  The lines used were H$\gamma ~\lambda 4340$, 
\ion{He}{1} $\lambda 4388$ and $\lambda 4471$, and  \ion{He}{2} $\lambda 4541$.  
Howarth et al. (1997) performed a cross-correlation
study similar to ours with {\it IUE} high resolution spectra of 
373 OB-type stars using the template star $\tau$~Sco = HD~149438 (B0.2 IV). 
Although their calibration technique differed slightly from ours, 
their findings supported those from our study.

\placefigure{f1}      
\setcounter{figure}{0}
\begin{figure}
\epsscale{0.8}
\plottwo{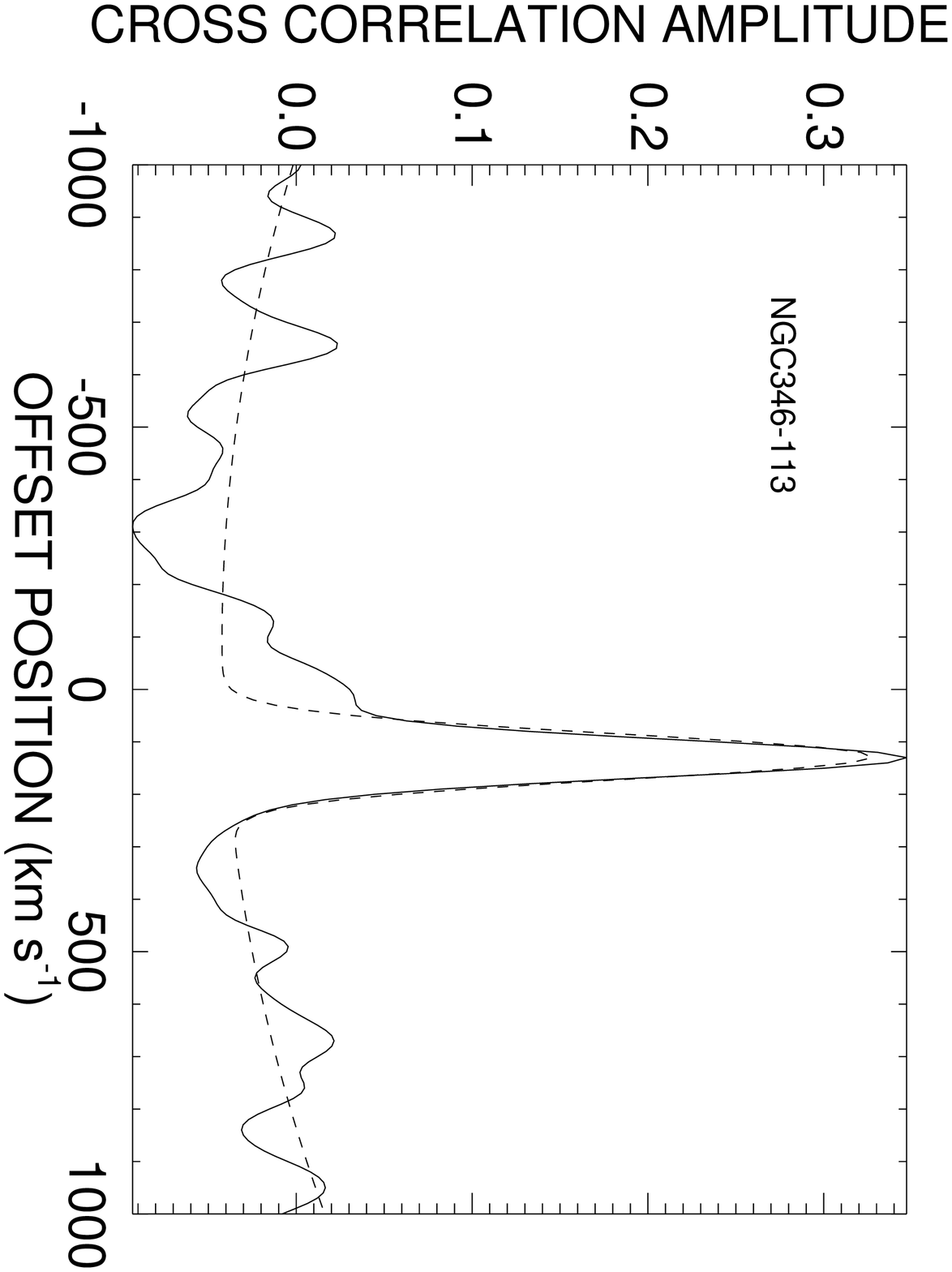}{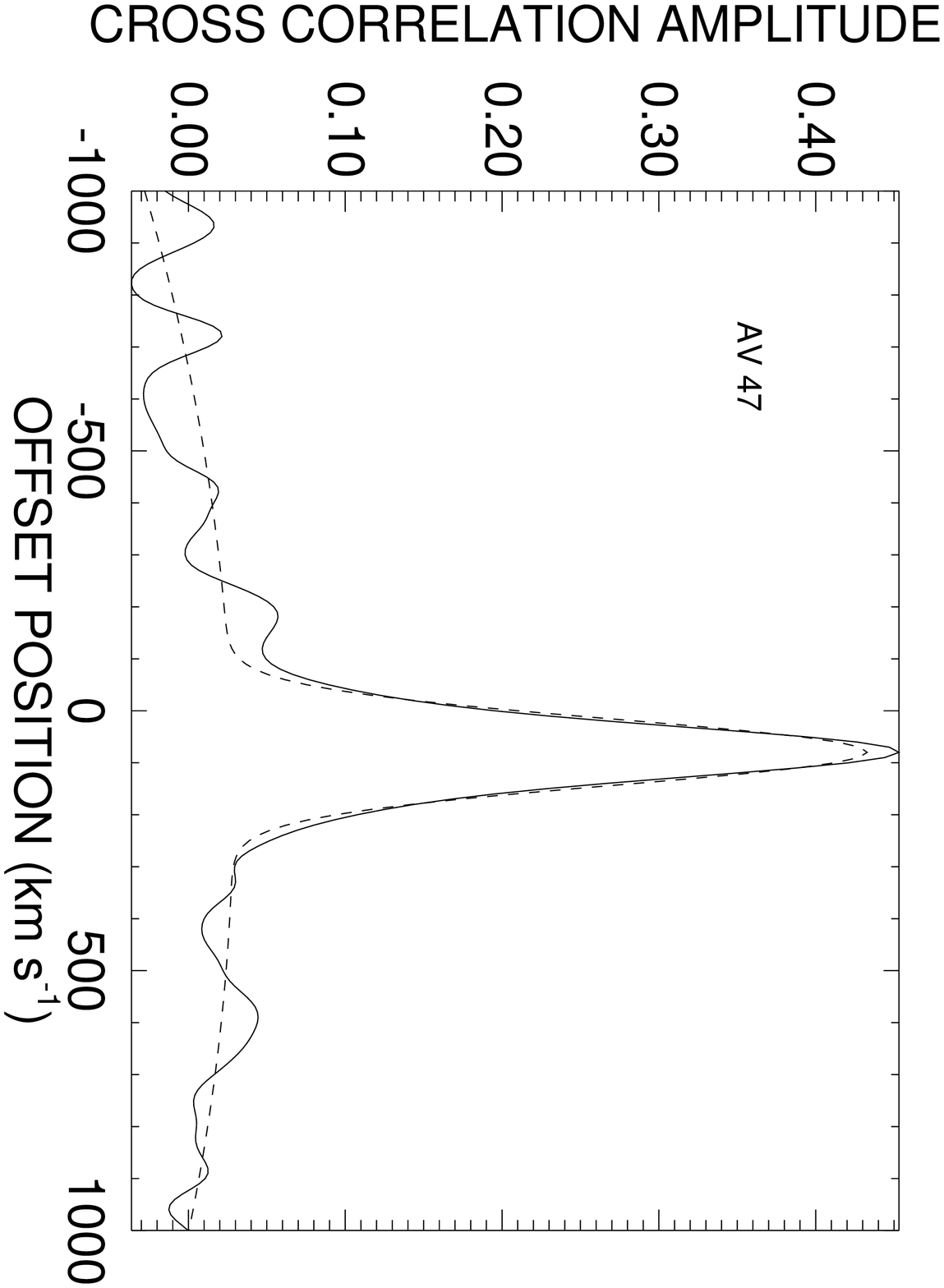}
\plottwo{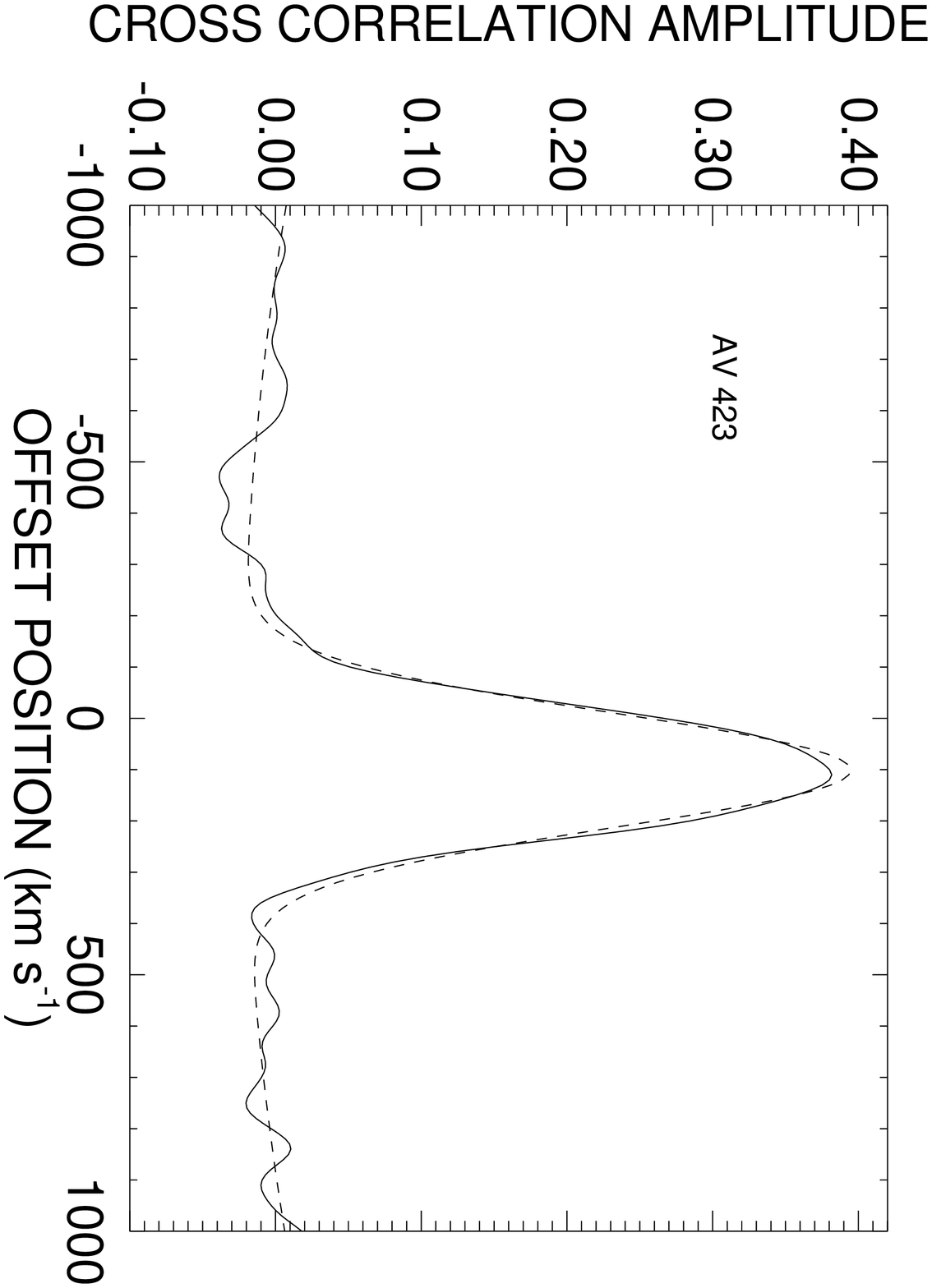}{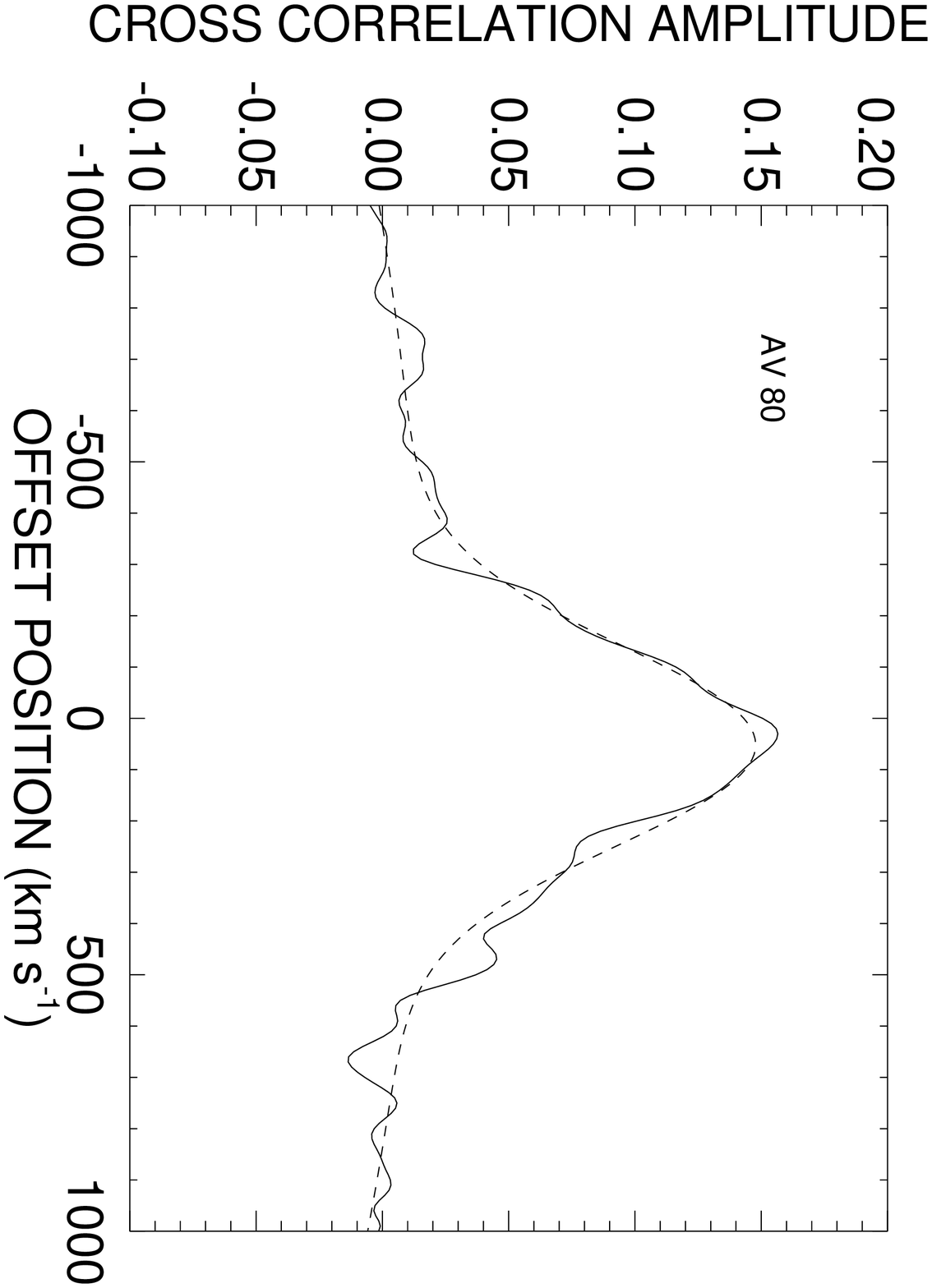}
\caption{The cross-correlation functions and Gaussian fits for the SMC stars 
(a) NGC346-113 (OC6~V; $V \sin i < 40$ km~s$^{-1}$), 
(b) AV~47 (O8~III((f)); $V \sin i = 76$ km~s$^{-1}$), 
(c) AV~423 (O9.5~V; $V \sin i = 186$ km~s$^{-1}$), (d) AV~80 (O4-6n(f)p; 
$V \sin i = 324 $km~s$^{-1}$).  The bumpy nature of the ccf for AV~80 may be 
indicative of non-radial pulsations.}
\label{f1}
\end{figure}

\section{Observations and Reductions}

Methods similar to those above can be utilized with observations 
of O-type stars made with the  Space Telescope Imaging
Spectrograph (STIS) E140M grating. 
The E140M grating produces in a single exposure an echelle spectrogram
covering the wavelength interval $1150 - 1700$\AA , with an effective 
spectral resolving power of $R = 46,000$.  This
grating set up is complementary to the {\it IUE} SWP arrangement which was 
used so effectively to obtain an almost complete catalog of UV
observations of Galactic O-type stars to $V \approx 11$. 
The Multimission Archive at Space Telescope 
(MAST\footnote{URL: http://archive.stsci.edu/}) 
contains spectra of the 18 SMC and 6 LMC O-type stars 
observed with this grating.  There are also STIS E140M grating
archival spectra of 12 Galactic O-type stars which we obtained 
in order to calibrate our methodology with the 
STIS observations.  Although the E140M spectra are similar in many
ways to the SWP high resolution spectra in our
previous methodology, it is prudent to use these overlapping
Galactic stars as an experimental checks of our
calibration. There are also available in MAST {\it IUE} SWP 
high resolution spectra of 4 SMC and 16 LMC O-type stars;
one of the latter is among those with archival STIS spectra.

The NEWSIPS MXHI files of the above SWP spectra are spectrograms 
that cover the wavelength region of $1200 - 1900$\AA . 
The archival STIS spectra are in extracted one dimensional FITS format, forming a near complete
spectrogram covering the wavelength range $1150 - 1700$\AA . 
We use a series of routines we have written in the
Interactive Data Language\footnote{IDL is a registered trademark of Research Systems, Inc.} 
to further reduce the spectra (see details in Penny, Gies, \& Bagnuolo 1997).  
One of the most important steps in
our reduction is the placement of the data on a $\log \lambda$ grid, 
in which each pixel step corresponds to a uniform
velocity step of 1 pixel = 10 km~s$^{-1}$.  The spectra were smoothed 
using a Gaussian transfer function (FWHM = 40 km~s$^{-1}$)
and the interstellar features are removed.  
The spectra are then rectified to a unit continuum. 

Complete lists of spectra obtained, targets, and their spectral classifications are
presented in Tables 1, 2, \& 3 for Galactic, SMC, and LMC targets, respectively. 
Spectral classifications given with square brackets are from investigators 
other than Walborn (see Ma\'{i}z-Apell\'{a}niz et al.\ 2004) and are taken
from the compilations of Howarth \& Prinja (1989), Conti \& Ebbets (1977), 
Garmany, Conti, \& Massey (1987), Conti, Garmany, \& Massey (1986), 
Parker et al.\ (1992), Garmany, Massey, \& Parker (1994), 
Massey et al.\ (1989), Massey, Parker, \& Garmany (1989), and Massey et al.\ (1995).  
We make special note of the SMC star AV~423. 
This star is classified as O9.5~V by Massey et al.\ (1995).  
However, its UV spectrum, plotted in Figure~2, 
is inconsistent with that designation.  
Both \ion{Si}{4}~$\lambda\lambda 1400$ and \ion{C}{4}~$\lambda\lambda 1550$ 
doublets have P~Cygni profiles, which makes this only the third star known in the SMC 
with \ion{Si}{4} emission.  This feature in SMC is indicative 
of a supergiant classification.  However, the \ion{N}{5} doublet is extremely weak 
even in absorption.  In addition to AV~423, we also plot 
AV~15 [O6.5 II(f)], AV~83 [O7 Iaf+], and NGC346-12 [O9.5-B0 V] for comparison in Figure~2.
Whatever the revised type of AV~423, it is clearly not a dwarf star, and we will treat it 
as an evolved object below. 

\begin{figure}
\epsscale{0.8}
\plotone{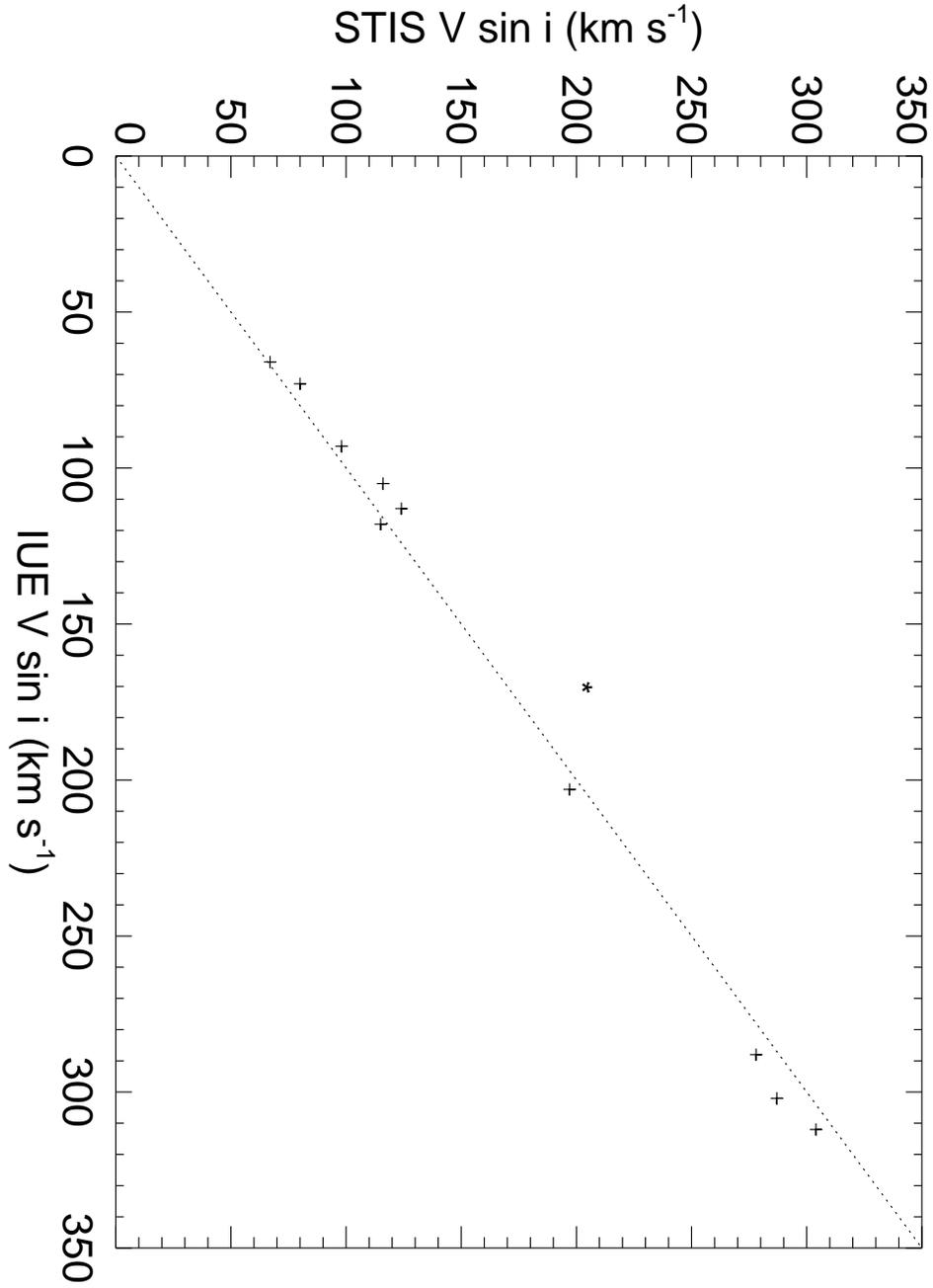}
\caption{Projected rotational velocities from STIS spectra versus those from 
{\it IUE} spectra for 12 Galactic O-type stars.  
The asterisk is plotted for the values for HD~13745, 
which may display time variations in line width (see \S4).}
\label{f2}
\end{figure}

\placefigure{f2}      
 
\section{Projected Rotational Velocities from Cross-Correlation Functions} 

The spectra for each star are cross-correlated with a narrow-lined template star.  
We actually cross-correlate each test star with
four different templates in order to get the best (usually highest) 
peak for fitting.  These templates are HD~34078 (O9.5 V; $V \sin i$ =  25 km~s$^{-1}$),  
HD~149438 (B0.2 IV; $V \sin i < 5$ km~s$^{-1}$), HD~54662 (O6.5 V; 
$V\sin i  =  85$ km~s$^{-1}$), and HD~57682 (O9 IV; $V \sin i  =  33$ km~s$^{-1}$). 
We cross-correlate essentially the full
wavelength range of the spectra (STIS: $1200 - 1700$\AA ; {\it IUE}: $1200 - 1900$\AA ), 
setting to unity regions surrounding the strong P~Cygni lines
of \ion{N}{5} $\lambda 1240$, \ion{Si}{4}  $\lambda$ 1400, and 
\ion{C}{4} $\lambda$ 1550, since these features reflect wind and not
photospheric rotational motion.  The region around Ly$\alpha$ is also set to unity.  
The spectra are padded with $1000$ km~s$^{-1}$
of artificial continuum on both ends to include the entire 
observed range in the cross-correlation.  The
cross-correlation function (ccf) is the sum of the square 
of the differences between the test spectrum and the reference
spectrum shifted in velocity from $-1000$ km~s$^{-1}$ to $+1000$ 
km~s$^{-1}$ at $10$ km~s$^{-1}$ intervals.  The ccfs are
rectified and inverted for convenience in fitting and plotting.  
Each ccf is then fitted with a Gaussian function, and $V \sin i$
values are derived from the Gaussian widths using the calibrations
 presented in P96 (HD~34078), Penny et al.\
(2001, HD~149438), Penny et al.\ (1997, HD~54662), and Penny (1997, HD~57682).  
We adopt 40 km~s$^{-1}$ as our minimum
measurable $V \sin i$, as this is the FWHM of the Gaussian transfer function used to smooth the data.

We compared $V \sin i$ values from our previous 
study to those obtained using  STIS  observations for those 
Galactic O-type stars in common.  This is especially important
since the wavelength range in the STIS spectra is smaller than that 
of the {\it IUE} observations, which might introduce
systematic differences in our $V \sin i$ values.
We plot in Figure~3 $V \sin i$ values from the STIS observations 
versus those from the {\it IUE} observations.  
We see excellent agreement, well within our estimated errors 
from the {\it IUE} data ($\pm 2-5\%$), with the
exception of HD~13745.  This star is noted in P96 as being a possible 
unresolved double-lined spectroscopic binary or it may be a short period
pulsator (Koen \& Eyer 2002), so a larger than expected difference in $V \sin i$
here is not unexpected. 
  
\placetable{t1}      
\begin{deluxetable}{lccccl}
\label{t1}
\tablewidth{0pc}
\tablecaption{Galactic Targets}
\tabletypesize{\footnotesize}
\tablehead{ 
\colhead{}  &  \colhead{} &  \colhead{$<V \sin i>$}  
 &  \colhead{$<V \sin i >$}  &  \colhead{} & \colhead{} \\ 
\colhead{} &\colhead{Spectral}  &    \colhead{(this paper)}  &
 \colhead{(P96)}   &  \colhead{SWP} & \colhead{STIS} \\
\colhead{Star} & \colhead{Classification} & \colhead{(km~s$^{-1}$)} & \colhead{(km~s$^{-1}$)} 
 & \colhead{\#} &\colhead{data set} 
}
\startdata 
HD~14434    &  O5.5 Vn((f))p &    377 &    368 &    16094 & O63508010 \\ 
HD~63005    & [O6 V]         & \phn80 & \phn73 &    21506 & O63531010 \\
HD~13268    & [O7]           &    287 &    302 & \phn9323 & O63506010 \\ 
HD~152590   & O7.5 V         & \phn67 & \phn66 &    16098 & O6LZ67010 \\ 
HD~12323    & ON9 V          &    124 &    113 & \phn9652 & O63505010 \\ 
HD~91651    & O9 V:n         &    278 &    288 &    14830 & O6LZ34010 \\ 
HD~156359   & [O9 III]       & \phn98 & \phn93 &    11274 & O6LZ70010 \\ 
HD~210809   & O9 Iab         &    115 &    118 & \phn9103 & O6359T010 \\ 
HD~308813   & O9.5 V         &    197 &    203 &    21172 & O63559010 \\ 
HD~168941   & O9.5 II-III    &    116 &    105 &    23843 & O6LZ81010 \\ 
HD~92554    & [O9.5 II]      &    304 &    312 &    16220 & O6LZ36010 \\ 
HD~13745    & O9.7 II((n))   &    196 &    168 &    16111 & O6LZ05010 \\
\enddata
\end{deluxetable}

\begin{figure}
\epsscale{0.8}
\plotone{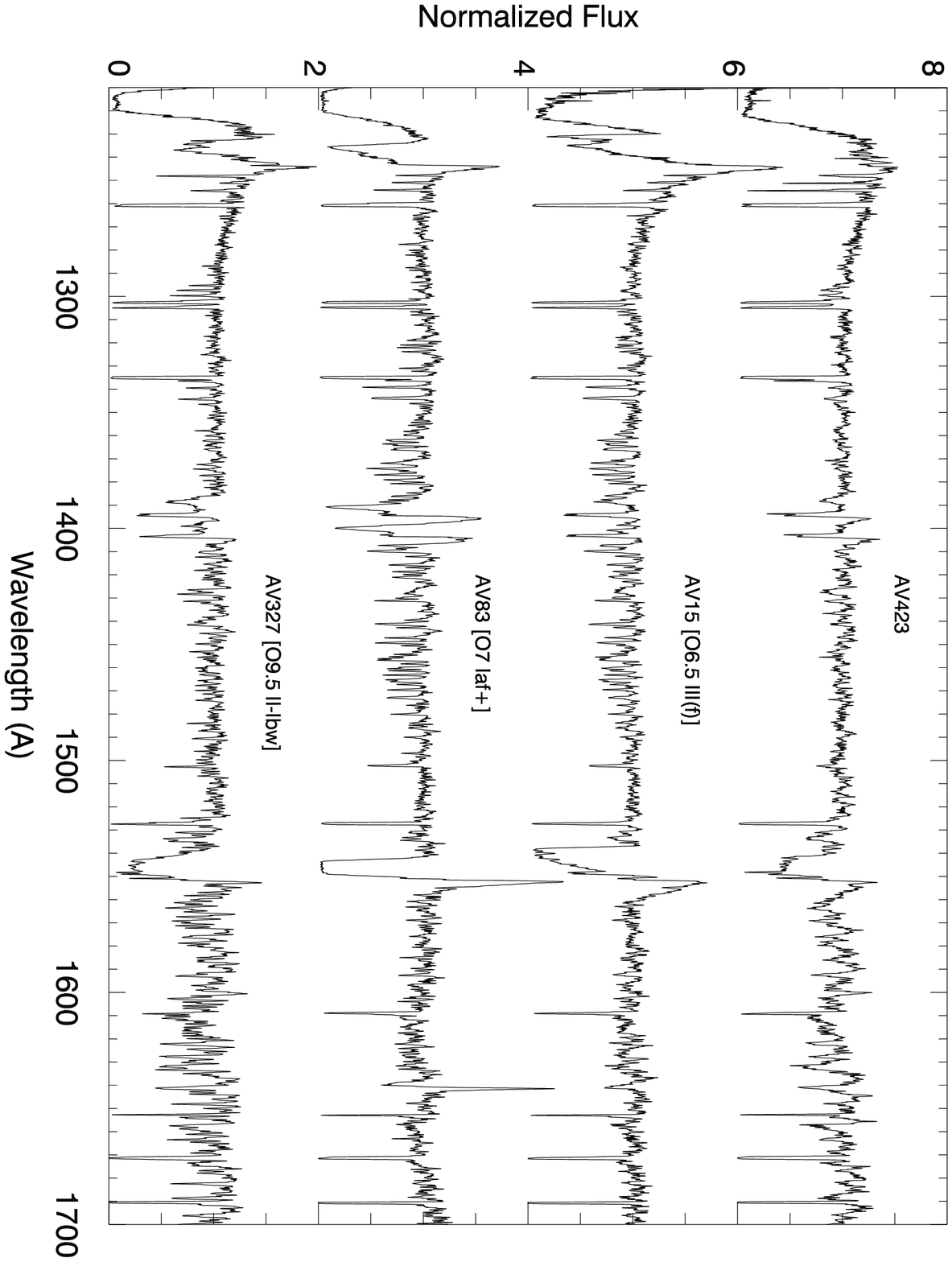}
\caption{STIS spectra of AV~423 ({\it offset by +6.0}), 
AV~15 (O6.5~II(f); {\it offset by +4.0}), AV~83 (O7~Iaf+; {\it offset 
by +2.0}), and AV~327 (O9.5~II-Ibw).  The current spectral classification 
of O9.5~V for AV~423 is inconsistent with its UV spectrum.}
\label{f3}
\end{figure}

\placefigure{f3}     
 
Our calculated projected rotational velocities are presented in Tables 2 and 3.  
Previous estimates of $V \sin i$ for the SMC stars in common 
by Walborn et al. (2000) are also shown in Table~2.  We note that they describe 
their values as preliminary due to the blended nature 
of the \ion{He}{1} + \ion{He}{2} $\lambda 4026$ line from which their 
estimates are determined.   Nevertheless, there is reasonable agreement 
between our results except in the cases of NGC346-324 and AV~47, 
which may be spectroscopic binaries observed at differing orbital phases. 
The SMC contains both the slowest and fastest rotators among 
our sample of Magellanic Cloud stars.  Both NGC346-113 and AV~220 have
narrow-lined spectra indicating a $V\sin i$ below the minimum for 
which we can determine an accurate projected rotational velocity ($<40$ km~s$^{-1}$). 
The fastest rotator is AV~80 that has a $V\sin i$ comparable to that 
found in some Be stars.  Walborn et al.\ (2000) show that the 
\ion{He}{2} $\lambda 4686$ feature is a double-peaked emission line 
in this star, which suggests that this star may have a disk-like 
circumstellar envelope, again like those found in Be stars. 

\placetable{t2}      
\begin{deluxetable}{lcrcl} 
\label{t2}
\tabletypesize{\scriptsize}
\tablewidth{0pc}
\tablecaption{Small Magellanic Cloud Targets}
\tablehead{ 
\colhead{}  &  \colhead{} 
 & \colhead{$<V \sin i>$}   &  \colhead{$<V \sin i>$}
 & \colhead{} \\ 
\colhead{}  &  \colhead{Spectral}
 & \colhead{(this paper)}   &  \colhead{(W00)}
 & \colhead{SWP \# or} \\
\colhead{Star} & \colhead{Classification}
 & \colhead{(km~s$^{-1}$)}  & \colhead{(km~s$^{-1}$)}
 & \colhead{STIS data set}   
}
\startdata 
NGC346-355    &  O2 III(f*)     & 112~~~~~~  &    120  & O4WR01010 \\ 
AV~388        &  [O4 V]         & 179~~~~~~  & \nodata & SWP33968 \\ 
NGC346-324    &  O4 V((f))      &  70~~~~~~  &    120  & O4WR01020 \\ 
NGC346-368    &  O4-5 V((f))    &  55~~~~~~  & \phn80  & O4WR01030 \\ 
AV~80         &  O4-6n(f)p      & 324~~~~~~  &    270  & O4WR12010,12020 \\ 
AV~75         &  O5 III(f+)     & 109~~~~~~  &    110  & O4WR11010,11020 \\
NGC346-113    &  OC6 Vz         &$<40$~~~~~~ & \phn45  & O4WR02010 \\ 
AV~243        &  [O6 III]       &  62~~~~~~  & \nodata & SWP33961 \\ 
NGC346-487    &  [O6.5 V]       &  65~~~~~~  & \nodata & O4WR23010 \\ 
AV~15         &  O6.5 II(f)     & 128~~~~~~  &    132  & O4WR14010,14020 \\ 
AV~220        &  O6.5f?p        &$<40$~~~~~~ & \phn62  & O4WR20010,20020 \\ 
SK80          &  [O6.5 If]      & 106~~~~~~  & \nodata & SWP06564,25438 \\ 
AV~95         &  O7 III((f))    &  82~~~~~~  & \phn89  & O4WR17010,17020 \\ 
AV~26         &  [O7 III]       & 127~~~~~~  & \nodata & SWP32466 \\ 
AV~83         &  O7 Iaf+        &  82~~~~~~  & \phn80  & O4WR15010,15020 \\ 
AV~69         &  OC7.5 III((f)) & 100~~~~~~  &    104  & O4WR13010,13020 \\ 
NGC346-682    &  [O8 V]         &  71~~~~~~  & \nodata & O4WR24010 \\ 
AV~47         &  O8 III((f))    &  76~~~~~~  &    169  & O4WR16010,16020 \\ 
AV~423        &  [O9.5 V](?)    & 186~~~~~~  & \nodata & O4WR18010,18020 \\ 
NGC346-12     &  O9.5-B0 V      &  67~~~~~~  & \phn84  & O4WR21010 \\ 
AV~327        &  O9.5 II-Ibw    &  71~~~~~~  & \phn80  & O4WR30010,30020 \\ 
AV~170        &  O9.7 III       &  54~~~~~~  & \phn65  & O4WR18010,18020 \\
\enddata
\end{deluxetable}

\placetable{t3}      
\begin{deluxetable}{lccl} 
\label{t3}
\tablewidth{0pc}
\tablecaption{Large Magellanic Cloud Targets}
\tabletypesize{\scriptsize}
\tablehead{ 
\colhead{}     & \colhead{Spectral} 
  &\colhead{$<V \sin i>$}    & \colhead{SWP \# or} \\ 
\colhead{Star} & \colhead{Classification}  
  & \colhead{(km~s$^{-1}$)}  & \colhead{STIS data set} 
}
\startdata 
HD~269810    & O2 III(f*)    &    173 & O6LZ11010 \\
SK~-66 172   & O2 III(F*)+OB & \phn68 & SWP33971 \\ 
HD~269698    & O4 I          &    157 & SWP08011,06967 \\
HD~269676    & O4-5 III(f)   &    125 & O63541010 \\
HD~269676    & O4-5 III(f)   &    156 & SWP05086,13908,14022 \\ 
SK~-69 212   & [O5 III(f)]   &    210 & SWP52709 \\ 
LH10-3120    & [O5.5 V((f*))]& \phn95 & O5EZ01010,01020,01030 \\ 
SK~-66 100   & [O6 III]      &    116 & SWP33139 \\
HD~270952    & O6 Iaf        & \phn61 & SWP45216 \\ 
HD~269357    & O6 I          & \phn97 & SWP04314,55231 \\ 
SK~-67 111   & [O6 Ia(n)fp]  &    209 & SWP10991,52745 \\ 
SK~-67 51    & [O6.5 III]    &    105 & SWP31341 \\
CD~-68 264   & [O8 V]        &    139 & SWP47851 \\ 
SK~-67 101   & [O8 II((f))]  &    101 & O4YN01010,01020 \\ 
LH58 52a     & [O8 II]       &    110 & SWP49323 \\ 
SK~-67 266   & [O8 Iaf]      &    144 & SWP20411 \\ 
LH58 19a     & [O9 II]       &   178  & SWP49312 \\ 
HD~268605    & O9.7 Ib       & \phn90 & SWP06540,51851,52010 \\ 
HD~269889    & O9.7 Ib       & \phn72 & SWP47601 \\ 
HD~269896    & ON9.7Ia       & \phn70 & SWP47594,55216 \\ 
SK~-67 106   & [B0 III]      &    135 & O4YN03010 \\ 
SK~-67 107   & [B0 III]      &    103 & O4YN04010,04020 \\
\enddata
\end{deluxetable}

\section{Results and Conclusions} 

While our ultimate goal is to determine the true rotational velocities of O-type stars
in differing environments, in general only the projected 
rotational velocities are measurable. However, we can make some
statistical arguments. The inclination angles of the polar axes $i$ 
of a group of stars should be randomly distributed. 
For each metallicity environment (Galactic, LMC, and SMC) we expect that 
the distribution of $i$ should be the same, that is random.  
Thus, we can safely intercompare the rotational properties 
through an examination of their projected rotational velocities. 
We can compare the projected rotational velocity distributions of the 
different groups of star by calculating their 
cumulative distribution function (cdf), as shown in Howarth et al.\ (1997) and 
Howarth \& Smith (2001).  The cdf gives the fraction of stars having a 
$V \sin i$ less than a specific upper value that ranges from zero to the 
maximum $V \sin i$ observed. 

Before comparing the MC samples it is helpful to determine if the 
cdf of Galactic dwarf (IV-V) stars is different from that of Galactic evolved (I-III) stars. 
We plot in Figure~4 the CDF for both groups of stars based on the sample in P96.  
We see that the evolved sample has both fewer slow rotators and fewer fast rotators 
compared to the unevolved sample. 
The Kolmolgorov-Smirnov (K-S) test determines the probability that two samples are drawn 
from the same parent population.  For the Galactic evolved and 
unevolved stars this probability is only $16\%$, which suggests that 
the differences may be significant.  Howarth et al.\ (1997) came to the 
same conclusion in their study of projected rotational velocities 
of Galactic O-stars.  They suggest that the evolved stars do indeed spin down 
as they grow in radius (removing the fastest rotators from the sample) and 
that there exists excess turbulent broadening among the supergiants (that 
removes the most narrow-lined stars from the sample).   

\begin{figure}
\epsscale{0.8}
\plotone{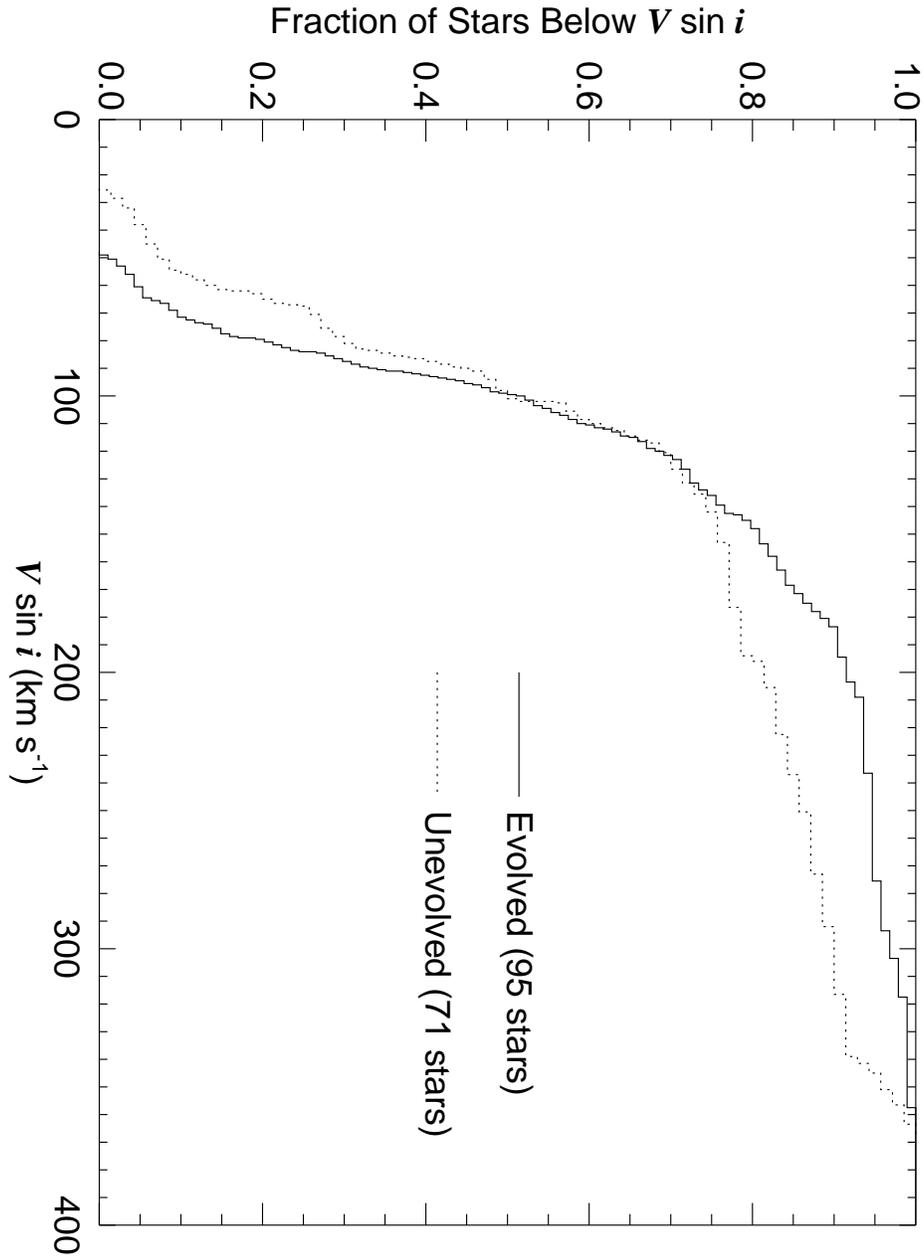}
\caption{The cumulative distributions of projected rotational velocities 
for Galactic stars from Penny (1996).  The dotted line 
represent those objects with luminosity classes IV and V, 
while the solid line corresponds to  
stars with luminosity classes I, II, and III.}
\label{f4}
\end{figure}

\placefigure{f4}     
 
The differences in the distributions of projected rotational velocity 
between the unevolved and evolved groups appears to support theoretical 
models that predict a spin-down during the core H-burning phases of 
evolution (Heger \& Langer 2000; Meynet \& Maeder 2000).  
It is important to recognize that most stars with an O spectral type 
are probably found in the core H-burning phase, even among the 
more luminous stars (see the evolutionary tracks of Schaller et al.\ 1992). 
Thus, we expect that the low $Z$, Magellanic Cloud O-stars are also 
in this phase where the rotational spin-down is predicted to be 
much less effective than we find in Galactic stars (Maeder \& Meynet 2001). 
The low $Z$ stars should retain an almost constant $V_{rot}$ 
during their core H-burning evolution. 
Therefore, we expected that the cumulative distribution function 
for the low $Z$ stars would fall well below that of Galactic stars, 
because of the relative excess of rapid rotators.  

The existence of a systematic difference between the velocity 
distributions of Galactic unevolved and evolved O-stars (Fig.~4) 
indicates that we must compare like samples between the 
Galactic and Magellanic Cloud stars in order to avoid spurious 
results because of differing proportions of unevolved and evolved stars in the groups. 
A large percentage of our Magellanic Cloud samples are objects with 
evolved luminosity classes (68\% and 90\% for the SMC and LMC samples, respectively).
Note that we include the two SMC stars lacking a luminosity class, AV~80 and AV~220,
among the evolved stars since Walborn et al.\ (2000) find them to have 
luminosities consistent with an advanced evolutionary state.  
We also include SMC star AV~423 among the evolved group because
of the features in the UV spectrum indicating a high luminosity (see \S 3).
We show in Figure 5 the cumulative distribution functions for the 
evolved O-stars in the Galaxy, SMC, and LMC.  Surprisingly, 
the distributions for all three samples are very similar, with K-S probabilities of
$44\%$ and $42\%$ that the LMC and SMC samples, respectively, are  
drawn from the same parent population as the Galactic sample. 
There is no visible trend of cumulative distribution functions with metallicity.  
The LMC sample has slightly faster $V\sin i$ values than that of the Galactic 
stars, but the SMC objects have slightly slower rotation rates than the other two samples.  
There are too few unevolved stars to make a reliable comparison, 
but the average projected rotational velocities in each group appear
to be comparable: $<V\sin i> = 131\pm 93$, $117\pm 31$, and $78\pm 45$ km~s$^{-1}$ 
for the Galactic, LMC, and SMC unevolved stars, respectively. 

\begin{figure}
\epsscale{0.8}
\plotone{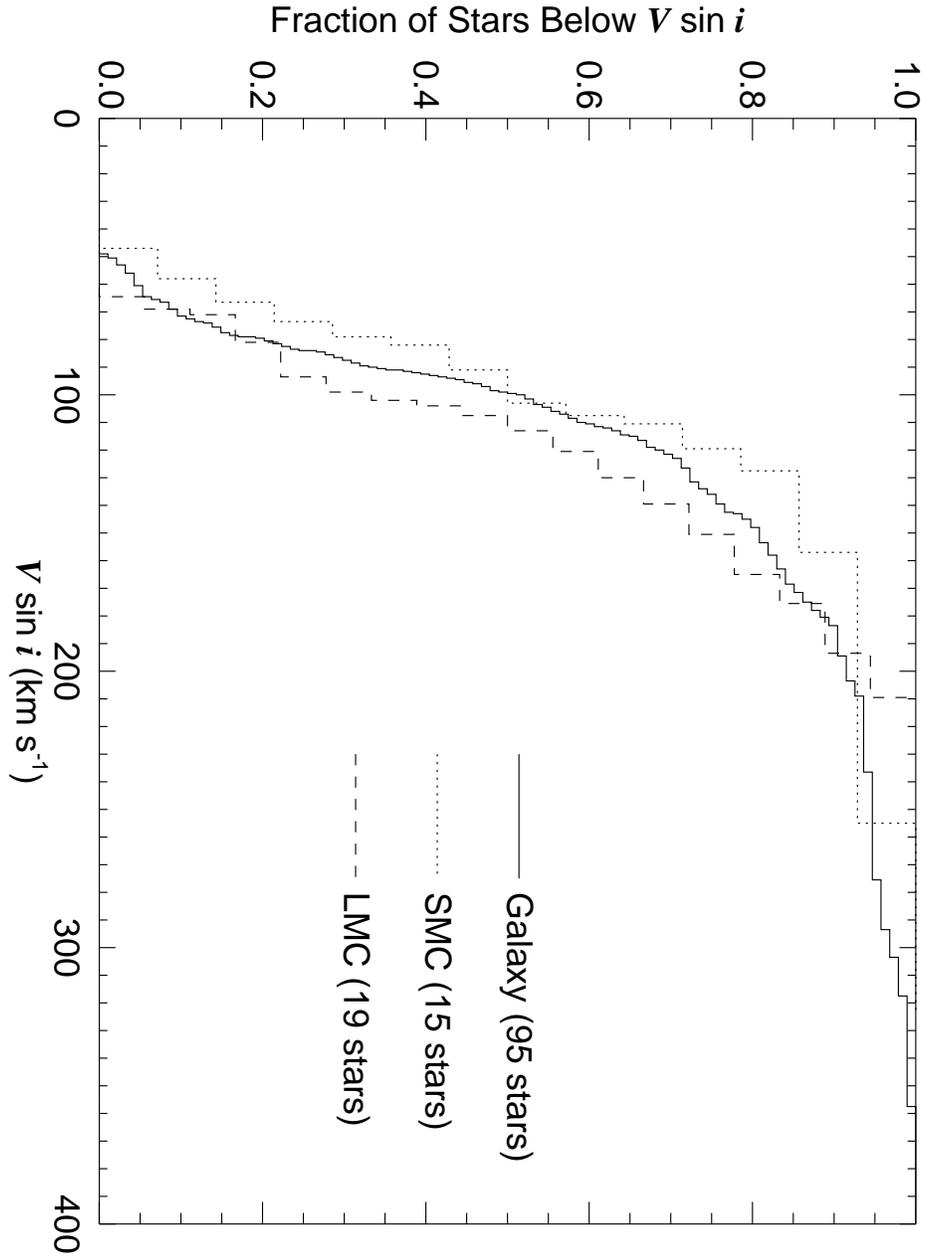}
\caption{The cumulative distribution of projected rotational velocities 
for 15 SMC ({\it dotted line}), 19 LMC ({\it dashed line}), and 
95 Galactic ({\it solid line}) O-type stars of luminosity class I to III.}
\label{f5}
\end{figure}

\placefigure{f5}     

Taken at face value, our results do not support the prediction that 
the low $Z$ stars of the Magellanic Clouds are rapid rotators.  
However, because of our relatively small sample sizes, we clearly 
need additional measurements to confirm the the result. 
Are low $Z$ stars actually rotating at the same speeds as 
their Galactic counterparts?  In order to determine conclusively the answer, 
a larger number of $V \sin i$ values for stars in these environments is needed. A previous
study by Keller (2001) examined the rotation rates of 100 early B-type stars in the LMC.  
Compared with a similar sample of Galactic stars, the low metallicity stars show larger
rotation rates.  The largest initial mass of these early B-type stars is only 12 $M_{\odot}$ 
and as such their MS evolution is not expected to be influenced by mass loss. However this
is exactly the sort of large scale project which is necessary for the more massive O-type stars.
A project with the FLAMES multi-object spectrograph on the VLT to obtain such spectra
is currently ongoing.
Additionally efforts are underway which focus on observing evolved (I-III classes) 
early type (O3-O5) stars in the LMC.  As these stars should correspond to a later 
stage in the core H-burning evolution of the most massive stars, they represent a 
critical test of the current treatment of mass and angular momentum loss 
in low metallicity massive stars.

\acknowledgments 

Support for this project AR\#09945 was provided in part by 
by NASA through a grant from the Space Telescope Science Institute, 
which is operated by the Association of Universities for Research in Astronomy, Inc. 
under NASA contract NAS5-26555.  
The data presented in this paper were obtained from the 
Multimission Archive at the Space Telescope Science Institute
(MAST).  Support for MAST for non-HST data is provided by the 
NASA Office of Space Science via grant NAG5-7584 
and by other grants and contracts.  


\clearpage






\end{document}